\documentclass[
    twocolumn,
	prd,
	amssymb,
	preprintnumbers,
	secnumarabic,
	nofootinbib,
	superscriptaddress]{revtex4-1}

\pdfoutput=1

\usepackage{graphicx}
\usepackage{enumitem}
\usepackage{latexsym}
\usepackage{amsfonts}
\usepackage{amssymb}
\usepackage{color}
\usepackage{xcolor}
\usepackage{amsmath}
\usepackage{slashed}
\usepackage{dcolumn}
\usepackage{verbatim}
\usepackage{float}
\usepackage{multirow}
\usepackage{xspace}
\usepackage[normalem]{ulem}
\usepackage{hyperref}

\definecolor{hypershade}{rgb}{0.8,0.3,0.3}
\hypersetup{
  pdfauthor={Nirmal Raj},
  pdftitle={Smoke and mirrors},
  pdfsubject={Mirror neutrons},
  colorlinks=true,
  citecolor=magenta,
  urlcolor=blue,
  linkcolor=hypershade
}

\newcommand{\gsim}{\gtrsim}
\newcommand{\lsim}{\lesssim}

\def\Oc{\mathcal{O}}

\newcommand{\acro}[1]{\textsc{\MakeLowercase{#1}}} %FOR `NOT SHOUTING' CAPS (e.g. acronyms)

\renewcommand{\tilde}{\widetilde} 
\newcommand{\beq}{\begin{equation}}
\newcommand{\eeq}{\end{equation}}
\newcommand{\bea}{\begin{eqnarray}}
\newcommand{\eea}{\end{eqnarray}}

\def\MNS{M_{\rm NS}}
\def\RNS{R_{\rm NS}}

\def\taunn{\tau_{nn'}}
\def\DelE{\Delta E} 

\allowdisplaybreaks

\begin{document}

\title{Smoke and mirrors: Neutron star internal heating constraints on mirror matter}

\author{David McKeen}
\email{mckeen@triumf.ca}
\affiliation{TRIUMF, 4004 Wesbrook Mall, Vancouver, BC V6T 2A3, Canada}

\author{Maxim Pospelov}
\email{pospelov@umn.edu}
\affiliation{School of Physics and Astronomy, University of Minnesota, Minneapolis, MN 55455, USA}
\affiliation{William I. Fine Theoretical Physics Institute, School of Physics and Astronomy, University of Minnesota, Minneapolis, MN 55455, USA}

\author{Nirmal Raj}
\email{nraj@triumf.ca}
\affiliation{TRIUMF, 4004 Wesbrook Mall, Vancouver, BC V6T 2A3, Canada}

\date{\today}

\begin{abstract}
Mirror sectors have been proposed to address the problems of dark matter, baryogenesis, and the neutron lifetime anomaly.
In this work we study a new, powerful probe of mirror neutrons: neutron star temperatures.
When neutrons in the neutron star core convert to mirror neutrons during collisions, the vacancies left behind in the nucleon Fermi seas are refilled by more energetic nucleons, releasing immense amounts of heat in the process.
We derive a new constraint on the allowed strength of neutron--mirror-neutron mixing from observations of the coldest (sub-40,000 Kelvin) neutron star, PSR 2144$-$3933.
Our limits compete with laboratory searches for neutron--mirror-neutron transitions but apply to a range of mass splittings between the neutron and mirror neutron that is 19 orders of magnitude larger.
This heating mechanism, also pertinent to other neutron disappearance channels such as exotic neutron decay, provides a compelling physics target for upcoming ultraviolet, optical and infrared telescopes to study thermal emissions of cold neutron stars.
\end{abstract}

\maketitle

{\bf \em Introduction.}---
Mirror world scenarios are one of the earliest ideas of ``dark sectors,'' i.e. a set of particles very weakly coupled to the Standard Model (SM) and neutral under SM gauge forces. They were originally motivated to restore parity symmetry by introducing states duplicate to the Standard Model (SM) charged under a new gauge group~\cite{mirrorintro:Kobzarev:1966qya,*mirrorintro:Foot:1991bp,*mirrorintro:Foot:1991py}, 
and more recently entertained in ``Twin Higgs'' solutions to the gauge hierarchy problem~\cite{twinhiggs:Chacko:2005pe}. Mirror states may provide a plethora of new effects: they could constitute cosmological dark matter~\cite{dm:asymreheat:Berezhiani:1995am,dm:PhysRevD.47.456,*dm:Berezhiani:2000gw,*dm:Berezhiani:2005vv,*twinhiggs:Garcia:2015loa,*twinhiggs:Garcia:2015toa},
address the cosmological asymmetry of baryons~\cite{baryog:PhysRevLett.87.231304,*baryog:Berezhiani:2008zza},
relax the GZK limit of cosmic rays~\cite{GZK:Berezhiani:2005hv,GZK:Berezhiani:2006je},
and collect in hidden stellar structures~\cite{mirrorstars:Blinnikov:1996fm,
*mirrorstars:Mohapatra:1999ih,
*mirrorstars:Foot:1999hm,
*mirrorstars:Foot:2001ne,
*mirrorstars:Curtin:2019lhm,
*mirrorstars:Curtin:2019ngc,
*mirrorstars:Hippert:2021fch}.
For reviews, see, e.g. Ref.~\cite{review:Okun:2006eb,*review:Foot:2014mia,*review:Berezhiani:2018zvs}.

A $\mathbb Z_2$ exchange symmetry connects the mirror sector to the SM. If this symmetry is respected by the vacuum expectation values of the scalars in the two sectors, the mirror states are expected to be exactly mass-degenerate with SM states.
This can be problematic for precision cosmology if the light degrees of freedom in the mirror sector carry a comparable amount of energy to ordinary photons and neutrinos.  This can be avoided if the $\mathbb Z_2$ is broken and the two sectors are asymmetrically reheated, $T_{\rm SM} >T_{\rm mirror}$~\cite{dm:asymreheat:Berezhiani:1995am}.
In such a case, nothing forbids a tiny mixing between the two sectors, leading to the possibility of transition between neutral SM states and those in the mirror sector. In this paper, we will address one of the most consequential of such mixings: that between the neutron, $n$ and the mirror neutron, $n^\prime$, which can lead to $n\to n^\prime$ transitions.

The $n$-$n^\prime$ system can be simply studied using a two-state Hamiltonian (with $c=1$),
\beq
\label{eq:H}
H = \begin{pmatrix} 
  m_n + \Delta E & \epsilon_{nn^\prime} \\
  \epsilon_{nn^\prime} & m_{n^\prime}
 \end{pmatrix},
\eeq
where $m_n\simeq m_{n^\prime}$ are the $n$ and $n^\prime$ masses, $\epsilon_{nn^\prime}$ is the mixing amplitude, and $\Delta E$ is the (medium-dependent) energy splitting between the two states. $n\to n^\prime$ transitions occur with a rate determined by $\epsilon_{nn^\prime}$ and $\Delta E$ and have been searched for using ultracold neutrons (UCN)~\cite{PSI:Ban:2007tp,*PSIlike:Serebrov:2009zz,*PSI:Altarev:2009tg,*Berezhiani:globalexpt:2017jkn,PSI:Kirch:2020kdg} and cold neutron beams~\cite{GZK:Berezhiani:2005hv}.
Assuming $n$-$n'$ mass degeneracy and a vanishing mirror magnetic field, $B^\prime=0$, the splitting $\Delta E$ can be neglected in such experiments and the transition probability after a time $t$ can be written $P_{n\to n^\prime}=(\epsilon_{nn^\prime} t)^2\equiv (t/\tau_{nn^\prime})^2$ (setting $\hbar=1$). 
The strongest constraint on this transition is $\tau_{nn^\prime}>352~\rm s$ or $\epsilon_{nn^\prime}<1.87\times10^{-18}~\rm eV$ at 95\%\,C.\,L.~\cite{PSI:Kirch:2020kdg}. 
Non-vanishing values of $m_n-m_{n^\prime}$ or $\Delta E$ quench $n\to n^\prime$ transitions and the limits from UCN experiments practically disappear for $\left|m_n-m_{n^\prime}\right|\gtrsim 10^{-12}~\rm eV$.

While $n \to n^\prime$ conversions are kinematically forbidden in stable nuclei due to nuclear binding energy [unless $m_n-m_{n^\prime}\gtrsim{\cal O}(\rm MeV)$] they can proceed in a neutron star (NS). 
In Ref.~\cite{NSmirror:Goldman:2019dbq} a limit of $\epsilon_{nn'} \lsim 10^{-13}$~eV was obtained from the change in the spin period of a NS induced by the production of mirror neutrons in the NS core.

In this {\em Letter} we present 
a new and powerful probe of mirror neutrons, namely NS temperatures.
When $n \to n'$ oscillations occur in inter-nucleon encounters in NS cores, tremendous heat is released: both the converted neutron and nucleon on which it scattered leave behind holes in the Fermi sea, which are refilled by higher-energy nucleons. 
This is analogous to the Auger effect in which energy is released as electrons cascade from higher energy atomic levels to unfilled lower energy ones. We find that luminosity measurements of the coldest pulsar observed constrain 
\begin{equation}
\epsilon_{nn'} \lsim 10^{-17}~{\rm eV},~{\rm or}~\tau_{nn'} \gsim 10{\rm~s},
\end{equation}
and that future measurements with ultraviolet, optical and infrared telescopes could further improve sensitivity, perhaps down to $\epsilon_{nn'} \simeq 10^{-20}$~eV ($\tau_{nn'} \simeq 10^5$~s).
In the NS system $\Delta E \simeq 10-100~\rm MeV$ due to nuclear interactions, thus our limits apply to $\left|m_n-m_{n^\prime}\right|\lesssim{\cal O}(10~\rm MeV)$, a much wider range of $m_{n^\prime}$ than that of terrestrial UCN searches.
We recently used this novel NS heating mechanism to set extensive limits on exotic neutron decay modes ($n \to$~BSM states)~\cite{NSheat:McKeenPospelovRaj2020}.
While in that study we had used a simplified picture of the NS to obtain order of magnitude estimates, in the present work we treat in more careful detail (i) the dynamics and constituents of the NS core and their implications for neutron disappearance, and (ii) astronomical measurements.
We also note that the mechanism of generating energy inside NSs via $n \to n'$ conversions was first pointed out in a pioneering work~\cite{NSmirror:Goldman:2019dbq} with application to pulsar binary observables. 
Here we update this calculation and apply it to what turns out to be a more sensitive observable: the NS luminosity. 

%%%%%%%%%%%%%%%%%
{\bf \em n-n$'$ conversion in NS.}---We start from the effective $n$-$n^\prime$ Hamiltonian in Eq.~(\ref{eq:H}).
In the NS medium, nuclear interactions induce values of $\Delta E$ that range from $\sim 10$ to $100~\rm MeV$~\cite{glendenning2000compact,McKeenNelsonReddyZhouNS}, and we can therefore ignore the ${\cal O}(\rm keV)$ splitting arising from the NS's magnetic field.
We also set $m_n=m_{n^\prime}$, noting that this does not materially change our results so long as $\left|m_n-m_{n^\prime}\right|\lesssim{\cal O}(10~\rm MeV)$.

NS cores contain large numbers of degenerate neutrons with a roughly 10\% admixture of protons in beta equilibrium. Therefore, mirror neutrons can be created in the scattering of neutrons on neutrons and protons, $nn\to n^\prime n$ and $np\to n^\prime p$. Due to the extreme degeneracy of nucleons in the NS core, Pauli blocking must be taken into account when computing the rates for these processes, as done in e.g. Ref.~\cite{Reddy:1997yr,*BertoniNelsonReddy}.
In the part of parameter space we will consider, the rate of mirror neutron production is small compared to the inverse lifetime of neutron stars. This means that the $n^\prime$ number density is not appreciable compared to that of nucleons and we do not have to consider their Pauli blocking. The rate of $n^\prime$ production is therefore written
%%%%%
\beq
\label{eq:prodrate}
\Gamma_{n^\prime}=\sum_{N=n,p}f_N n_N\langle\sigma_{n^\prime N}v\rangle_{p_N>p_{F_{N}}}~,
\eeq
%%%%%
where $n_N$ is the nucleon number density and $f_n=1/2$, $f_p=1$ takes into account the scattering of identical states or not. The cross section to produce a mirror neutron through scattering on a nucleon, $\sigma_{nN\to n^\prime N}$, is averaged over the initial particle momenta, taking their relative velocity $v$ into account, as well as the Pauli blocking of the final state nucleon as indicated by the subscript $p_N>p_{F_{N}}$.
To do this, we take the initial nucleon momentum distributions to be those of non-interacting fermions at zero temperature,
%%%%%%
\beq
\label{eq:momdist}
f_N(p)=\Theta\left(p_{F_{N}}-p\right),
\eeq
%%%%%%
where $p_{F_{N}}$ is the Fermi momentum of the nucleon $N$. 
The nucleon number density is related to $p_{F_{N}}$ through
$n_N=p_{F_{N}}^3/(3\pi^2)$.

We relate the $n^\prime$ production cross sections to the elastic neutron scattering cross sections:
\begin{equation}
\sigma_{n^\prime N} \simeq  g_N \left(\frac{\epsilon_{nn'}}{\Delta E}\right)^2\sigma_{nN\to nN}~,
\end{equation}
with $g_n=2$, $g_p=1$ a multiplicity factor (note that $f_ng_n=f_pg_p=1$) and $\epsilon_{nn'}/\Delta E$ an effective in-medium $n$-$n^\prime$ mixing angle.

We obtain neutron-nucleon cross sections using experimentally determined phase shifts~\cite{NNOnline}.
We conservatively use only the $s$-wave phase shifts, noting that higher order partial waves are subdominant. The cross sections can be written in terms of the spin singlet and triplet phase shifts, $\delta_S$ and $\delta_T$ respectively, as
%%%%%%
\begin{align}
 \sigma_{nn\to nn}&\simeq\frac{1}{4}\times\frac{16\pi}{m_N^2v^2}\sin^2\delta_S,
\\
\sigma_{np\to np}&\simeq\frac{1}{4}\times\frac{16\pi}{m_N^2v^2}\left(\sin^2\delta_S+3\sin^2\delta_T\right)~,
\label{eq:xs}
\end{align}
%%%%%%
where the factors of $1/4$ come from spin-averaging.
In the energy region of interest $\sigma_{np} \gtrsim 4 \sigma_{nn}$.

Finally, we perform the final state phase space integration taking into account the nucleon momentum distribution in Eq.~(\ref{eq:momdist}) and only include final state nucleon momenta above the Fermi momentum as indicated in the subscript of Eq.~(\ref{eq:prodrate}). 
With all of these taken into account, for total nucleon density $n_{\rm nuc}$ (with $n_p \simeq 0.1 n_n$) we find $n^\prime$ production rates of 
%%%%%
\beq
\Gamma_{n^\prime}=\frac{1}{1.2\times 10^{11}~\rm yr}\left(\frac{\epsilon_{nn'}}{10^{-17}~{\rm eV}}\right)^2\left(\frac{n_{\rm nuc}}{0.3~{\rm fm}^{-3}}\right)~.
\label{eq:prodrateexample}
\eeq
%%%%%
The observable in which we are interested is the NS temperature, related to the energy liberated in the $n\to n^\prime$ conversions described above.
In any given conversion, there are three sources of energy transmitted to the rest of the star:
(i) the energy from the filling of the hole left by the converted neutron, $\tilde\mu_{n}-E_n^i$,
(ii) that from filling the hole left by the spectator nucleon, $\tilde\mu_{N}-E_N^i$, and
(iii) the excess kinetic energy of the final state nucleon kicked up above the Fermi sea, $E_N^f-\tilde\mu_{N}$.
Here $\tilde\mu_{N}$ refers to the kinetic energy of a nucleon at the top of the Fermi sea, which is {\em not} $p_{F_N}^2/2m_N$ when the potential energy of strong interactions is accounted for, and the $i$ and $f$ superscripts refer to initial or final state particles.
Putting all of this together, the energy released in any given conversion is therefore $\tilde\mu_{n}-p_{n^\prime}^2/2m_{n^\prime}$ and the total energy injection rate is:
%%%%%
\beq
\label{eq:Erate}
\dot E_{n^\prime} = \sum_{N=n,p}f_N n_N\left\langle\left(\tilde\mu_{n}-\frac{p_{n^\prime}^2}{2m_{n^\prime}}\right)\sigma_{n^\prime N}v\right\rangle_{p_N>p_{F_{N}}}~.
\eeq
%%%%%
We implicitly take scattering at these energies to be elastic, a conservative assumption as it releases less energy than inelastic scattering.

{\bf \em Neutron star heating.}---
The total energy released per unit time by $n\to n^\prime$ conversions in the NS is
%%%%%
\begin{equation}
    L_{n\to n^\prime} = \int d^3r\, n_n({\bf r})\dot E_{n^\prime}({\bf r})~.
    \label{eq:lumi}
\end{equation}
%%%%
To compute this quantity, we assume spherical symmetry and use high-density nuclear equations of state (EoS) from the literature to solve the Tolman–Oppenheimer–Volkoff (TOV) equations that govern the hydrodynamic stability of the star~\cite{TOV:Tolman:1939jz,*TOV:Oppenheimer:1939ne}.
From this, we can use the resulting $n_N=p_{F_N}^3/3\pi^2$, $\tilde\mu_{N}$, and $\Delta E$ (which we approximate as the average energy per nucleon as a proxy for the neutron self-energy~\cite{glendenning2000compact,McKeenNelsonReddyZhouNS}) to compute $\dot E_{n^\prime}$ as a function of radial co-ordinate $r$.
We can then ask whether the heating rate computed in Eq.~(\ref{eq:lumi}) for a given NS is compatible with its temperature, $T_{\rm NS}$. 
Old, cold NSs of age $\gsim 10^5$~yr cool mainly via photon emission from their surfaces and have approximately uniform temperatures~\cite{NSCooling:Yakovlev:2004iq}. 
Taking the luminosity to be that of a blackbody, $L_\gamma=4\pi R_{\rm NS}^2\sigma_{\rm SB}T_{\rm NS}^4$, where $\sigma_{\rm SB}$ is the Stefan-Boltzmann constant, we can set an upper bound on the transition amplitude $\epsilon_{nn'}$ by requiring that $L_{n\to n'}\leq L_\gamma$. 

Solving the TOV equations with the `BSk24' EoS from Ref.~\cite{BSk:Pearson:2018tkr} for a central density that produces a NS with mass $M_{\rm NS}=1.5 M_\odot$ and radius $R_{\rm NS}=12.6~\rm km$, characteristic of a typical NS, and computing the luminosity in Eq.~(\ref{eq:lumi}), we arrive at the following upper bound on the transition amplitude applicable for an old, cold NS with surface temperature $T_{\rm NS}$,
%%%%%
\beq
\epsilon_{nn^\prime} < 1.6 \times 10^{-17}~{\rm eV} \bigg(\frac{T_{\rm NS}}{3\times 10^4~{\rm K}}\bigg)^2~.
\label{eq:epsboundestimate}
\eeq
%%%%%
We see that measurements or limits on temperatures of such NSs at the $10^4-10^5~\rm K$ level provide limits that are competitive with searches for transitions of UCN to mirror neutrons.
Our NS configuration corresponds to volume-averaged values of
$\langle \Delta E \rangle$ = 29 MeV (with $\Delta E \rangle$ = 79 MeV at the NS center and 2 MeV near the outer edge of the core),
$\langle \tilde\mu_{n} \rangle = 68$ MeV,
$\langle \tilde\mu_{p} \rangle = 13$ MeV,
$\langle n_n \rangle = 0.22$~fm$^{-3}$,
$\langle n_p \rangle = 0.02$~fm$^{-3}$. 
Remarkably, the proton-catalyzed $n^\prime$ production contribution to $L_{n\to n'}$ dominates over that from neutron-catalyzed production by a factor of $\sim 4$ for two reasons: (i) $n_p<n_n$ implies less Pauli blocking and (ii) $\sigma_{np}>\sigma_{nn}$ as mentioned above.

\begin{figure*}[t]
\centering
\hspace*{-.5in}
  \includegraphics[width=\textwidth]{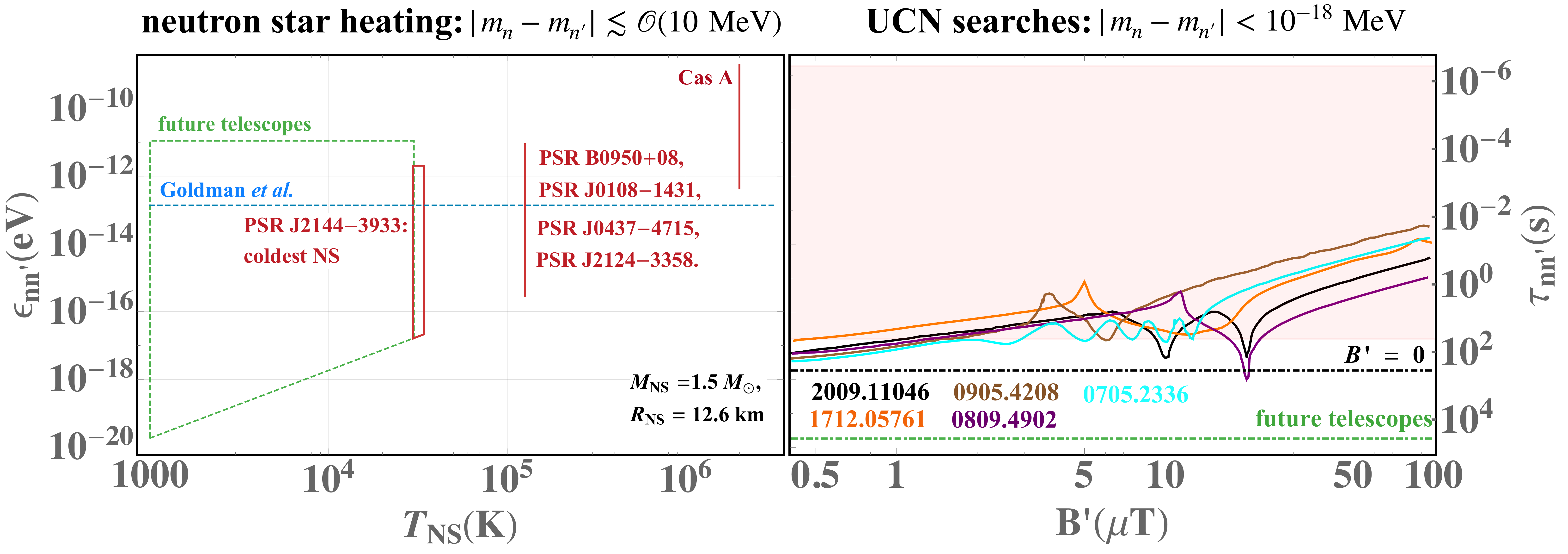}
  \caption{\textbf{\em Left.} Limits from neutron star temperatures on neutron-mirror mixing $\epsilon_{nn'}=1/\tau_{nn'}$, assuming the `BSk24'~\cite{BSk:Pearson:2018tkr} high-density equation of state giving an NS mass and radius of 1.5 $M_\odot$ and 12.6~km. 
  The strongest upper bound comes from the coldest NS observed, PSR J2144$-$3933, and we indicate the range in upper bounds on $\epsilon_{nn'}$ coming from the spread in upper bounds on its temperature. 
  The largest values of $\epsilon_{nn'}$ are probed by the youngest NS confirmed, Cas A. 
  Other pulsars consistent with a temperature of $10^{5.1}$~K also provide bounds in the intermediate region.
  Future ultraviolet, optical and infrared telescopes could measure NS temperatures down to 1000~K: the green region depicts the resulting sensitivity to $\epsilon_{nn'}$ for a given $T_{\rm NS}$. 
  Also shown is the bound obtained by Goldman {\em et al.} in Ref.~\cite{NSmirror:Goldman:2019dbq} from NS rotation periods.
  \textbf{\em Right.} Limits from terrestrial experiments looking for $nn'$ oscillations as a function of an unknown local mirror magnetic field strength $B'$, as adopted from Ref.~\cite{PSI:Kirch:2020kdg}; also shown is the limit for $B' = 0$.
  The arXiv numbers of the various experimental publications are indicated.
  These limits apply only in the extremely mass-degenerate limit and weaken considerably for $m_n' - m_n > 10^{-18}$ MeV, whereas our NS heating limits are valid so long as $m_n' - m_n < \DelE \simeq 10$~MeV. 
  For comparison, the region shaded red is excluded by our NS heating limits. See text for more details.} 
   \label{fig:limits}
\end{figure*}

The tightest upper bound on $\epsilon_{nn^\prime}$ comes from the coldest NS that has been observed, PSR J2144$-$3933~\cite{coldestNSHST}. As in Ref.~\cite{coldestNSHST}, we can determine an upper bound on this NS's temperature by modelling its spectral flux density as
%%%%%
\beq
f_\nu = \bigg(\frac{R_\infty}{d_\star}\bigg)^2 \frac{2 \pi h f^3_\nu}{c^2} \frac{10^{-1.24 c_{\rm ex}}}{\exp(h \nu/k_B T_\infty ) -1}~,
\eeq
%%%%%
where $T_\infty=T_{\rm NS}/(1+z)$, $R_\infty=R_{\rm NS}(1+z)$, with $(1+z)~=~(1-2 G \MNS/\RNS c^2)^{-1/2}$. 
(We have restored $c$ and $h=2\pi\hbar$ here.)
We then require that the average spectral flux density in the frequency range $1.51\times10^{15}\,{\rm Hz}<\nu<2.21\times10^{15}\,{\rm Hz}$ not exceed $5.9~\rm nJy$, for $157\,{\rm pc}<d_\star<192\,{\rm pc}$ and $0<c_{\rm ex}<0.06$. This translates into a range of upper bounds $T_{\rm NS}<29600~\rm K$ to $T_{\rm NS}<34100~\rm K$. The left panel of Fig.~\ref{fig:limits} shows the resulting upper limit on $\epsilon_{nn^\prime}$ from this range using Eq.~(\ref{eq:epsboundestimate}). 
%$$$$$$$$$
%$$$$$$$$$

The limit on $\epsilon_{nn^\prime}$ can change slightly if we consider different NS masses. In particular, larger NS masses lead to higher densities and Fermi momenta, reducing the $n\to n^\prime$ rate; using the same EoS as above, `BSk24', with $M_{\rm NS}=2.27M_\odot$ and $R_{\rm NS}=11~\rm km$ gives a 2.5 times weaker bound on $\epsilon_{nn^\prime}$. However, keeping the NS mass fixed to our nominal value, $M_{\rm NS}=1.5M_\odot$, and varying the EoS changes the bound by less than 10\% as we confirmed by comparing with results using `BSk22', `BSk25', and `BSk26'~\cite{BSk:Pearson:2018tkr}.

There is a ceiling to the limit we have derived above---a value above which our analysis no longer simply applies---occurring roughly at the value of $\epsilon_{nn^\prime}$ that results in a volume average of the $n^\prime$ production rate in Eq.~\eqref{eq:prodrateexample} that is equal to the inverse age of the NS. 
To extend our analysis to larger $\epsilon_{nn^\prime}$ would require considering a non-negligible $n^\prime$ number density in the NS.
The non-trivial phase space distribution of the mirror neutrons complicates the problem and we defer an analysis to future work~\cite{NS-MPR}.
The lifetime of PSR J2144$-$3933 is estimated to be $3\times 10^8~\rm yr$~\cite{coldestNSHST}, cutting off our exclusion at $\epsilon_{nn^\prime}=6.2\times10^{-12}~\rm eV$. 

In addition, the left panel of Fig.~\ref{fig:limits} shows similarly derived limits from the thermal luminosities of the NSs  
PSR B0950$+$08~\cite{NStemp:B0950+08},
PSR J0108$-$1431~\cite{NStemp:J0108-1431},
PSR J0437$-$4715~\cite{NStemp:J0437-4715}, and
PSR J2124$-$3358~\cite{NStemp:J2124-3358},
all of which are consistent with $T_{\rm NS} = 10^{5.1}~\rm K$.
The youngest of these is PSR B0950$+$08 with a spin-down age of $10^{7.2}~\rm yr$~\cite{NStemp:B0950+08}, setting the ceiling in this temperature range.

To probe larger values of $\epsilon_{nn^\prime}$ requires younger NSs. 
In contrast to older NSs, those younger than about $10^5$~yr cool through emission of neutrinos from the high density core, which scales differently with the core temperature than blackbody emission from the surface considered above.
The youngest observed NS is Cas A (CXOU J232327.8$+$584842) with an age of $340~\rm yr$. 
Its surface temperature is $2\times 10^6~\rm K$, corresponding to an internal temperature $T_{\rm int} = 3\times 10^8~\rm K$ assuming an Fe envelope~\cite{CasA}. 
Assuming a minimal cooling scenario through the modified Urca process~\cite{NSCooling:Page:2004fy,NSCooling:Yakovlev:2004iq}, its neutrino luminosity is $L_\nu\simeq 10^{32}\,{\rm erg/s}\,(T_{\rm int}/10^8~{\rm K})^8$. Requiring $L_{n\to n^\prime}<L_\nu$ results in the upper limit $\epsilon_{nn^\prime}<3\times10^{-13}~\rm eV$, also shown in the left panel of Fig.~\ref{fig:limits}. Given Cas A's age, this applies for $\epsilon_{nn^\prime}\lesssim 2\times 10^{-9}~\rm eV$. 
If we assume that the dust blob in the location of SN 1987A is powered by the thermal luminosity of a 34 yr-old NS~\cite{NS1987A}, we could access $\epsilon_{nn'} \lesssim 6\times 10^{-9}~\rm eV$.
The upper limit on $\epsilon_{nn'}$ from this potential NS, whose internal temperature is constrained to be $10^9~\rm K$~\cite{NS1987A}, is $\epsilon_{nn'} \lesssim 6 \times 10^{-11}$~eV assuming modified Urca neutrino cooling.

The green region in the left panel of Fig.~\ref{fig:limits} shows the future reach obtainable by measuring the thermal luminosity of PSR J2144$-$3933 (which should have reached a temperature of 1000 K at $t \simeq 2 \times 10^7$~yr and must now be much colder than the present upper bound on its temperature~\cite{NSCooling:Page:2004fy,NSCooling:Yakovlev:2004iq}) or other, yet undiscovered, cold NSs, by current and imminent ultraviolet, optical and infrared survey telescopes,
e.g. LUVOIR~\cite{LUVOIR},
Rubin/LSST~\cite{RubinLSST},
Dark Energy Survey~\cite{DES}, and
Roman/WFIRST~\cite{RomanWFIRST}.
It has also been proposed that NSs down to 1000~K temperatures could be observed in upcoming infrared telescopes over realistic integration times~\cite{NSvIR:Baryakhtar:DKHNS} (leading to studies of NSs heated to infrared temperatures via capture of ambient dark matter~\cite{NSvIR:Baryakhtar:DKHNS,NSvIR:Raj:DKHNSOps,*NSvIR:SelfIntDM,*NSvIR:Bell2018:Inelastic,*NSvIR:GaraniGenoliniHambye,*NSvIR:Queiroz:Spectroscopy,*NSvIR:Bell2019:Leptophilic,*NSvIR:Hamaguchi:Rotochemical,*NSvIR:GaraniHeeck:Muophilic,*NSvIR:Pasta,*NSvIR:Riverside:LeptophilicShort,*NSvIR:Marfatia:DarkBaryon,*NSvIR:Riverside:LeptophilicLong,*NSvIR:Bell:Improved,*NSvIR:DasguptaGuptaRay:LightMed,*NSvIR:GaraniGuptaRaj:Thermalizn,*NSvIR:Bell:ImprovedLepton,*NSvIR:Queiroz:BosonDM} and exotic neutron decay~\cite{NSheat:McKeenPospelovRaj2020}).
The ceiling here corresponds to $\Gamma^{-1}_{n^\prime} = 10^7$~yr, describing the age of an NS beyond which we expect it to have $\lsim \Oc(10^3)$~K temperatures~\cite{NSCooling:Page:2004fy,NSCooling:Yakovlev:2004iq}.

In addition, we show the upper bound on the transition amplitude that comes from the analysis of Ref.~\cite{NSmirror:Goldman:2019dbq} from the change in the spin period of NSs along with our limits in the left panel of Fig.~\ref{fig:limits}.

For comparison, in the right panel of Fig.~\ref{fig:limits} we show the limit on $\epsilon_{nn'} = \taunn^{-1}$ as a function of the local mirror magnetic field $B'$, from terrestrial $n$-$n'$ oscillation searches~\cite{PSI:Ban:2007tp,*PSIlike:Serebrov:2009zz,*PSI:Altarev:2009tg,*Berezhiani:globalexpt:2017jkn,PSI:Kirch:2020kdg}.
The dot-dashed grey line depicts the strongest limit for $B' = 0$, $\tau_{nn'} > 352$~s~\cite{PSI:Kirch:2020kdg}. While these limits are comparable to those that we have derived from NS heating, they only apply to the situation of extreme degeneracy, $|m_{n'} - m_n| < \mu_n B' \simeq 10^{-18}~$MeV, with the oscillation probability quenched for larger splittings. On the other hand our NS limits apply for mass splittings up to 19 orders of magnitude greater, $|m_{n'} - m_n| \lsim \Oc(10)$~MeV, as determined by the $n$-$n^\prime$ splitting in the NS medium.

%%%%%%%%%%%%
{\bf \em Discussion.}---
We have shown in this study that, via liberating heat from the energy stored in the Fermi sea of neutron stars, neutron oscillations to mirror neutrons could be probed in NS thermal emissions in wide regions of parameter space inaccessible to terrestrial searches.
A strong hint of $n \to n'$ conversions would be a universal minimum temperature detected over a statistical ensemble of NSs using current and imminent telescopes.
Whether such a signal can be definitely attributed to $n \to n'$ oscillations would depend much on the success of efforts to pinpoint NS masses and radii, and the high-density equation of state of nuclear matter such as undertaken by the NICER experiment~\cite{NICER1:Riley:2019yda,*NICER2:Raaijmakers:2019qny,*NICER3:Miller:2019cac}. 
If the EoS can be precisely constrained, uncertainties on our prediction for the NS luminosity in Eq.~\eqref{eq:lumi}, containing several EoS-dependent quantities, could be minimized.
(We note that if ``dark baryons" such as mirror neutrons are present in NSs in appreciable quantities, the EoS may be non-trivially impacted~\cite{PaoloNS:2008db,*PaoloNS:2010ji,McKeenNelsonReddyZhouNS,*SheltonNS,*MottaNS,*EllisPattavinaNS}.)

 Another scenario probed by NS heating is neutron decay to any exotic final state, as we had studied for dark baryons in Ref.~\cite{NSheat:McKeenPospelovRaj2020} and will explore in more detail in forthcoming work~\cite{NS-MPR}.
 Here NS heating is the sole probe of dark baryons for tiny mixings with neutrons.
 The need for astronomical observations of low-temperature neutron stars is usually motivated by the need to constrain cooling models and the EoS~\cite{NSLumi:Potekhin:2020ttj,NSCooling:Yakovlev:2004iq,NSCooling:Page:2004fy}, and to check for reheating mechanisms, e.g. rotochemical heating~\cite{Rotochemical1:Koichi} or possible accretion of interstellar material.
Our studies of $nn'$ oscillation and $n$ decay, probing the properties of neutrons directly in NSs, provide
further compelling motivations for such a campaign. 
We also note that precise measurements of cooling curves of young NSs would also be useful, as these would help set a sharper upper bound on the neutron mixing than presented here.
Precision measurements and fits to cooling models would also make our bounds stronger than the ones obtained via our loose criterion, $L_{n\to n'} \leq L_{\rm NS}$.

 We remark that NS heating will not be as competitive as terrestrial searches in probing the $|\Delta B| = 2$ process of neutron oscillations to anti-neutrons ($\bar n$).  
 This process would release at least $2 m_n \simeq$~2 GeV per $n \to \bar n$ conversion in the NS core via the $\bar n$ annihilating with surrounding neutrons.
 Keeping all other quantities fixed, the NS luminosity in Eq.~\eqref{eq:lumi} is now 10--100 times higher, implying that the limits on the the off-diagonal mass $\epsilon_{n \bar n}$ are 3--10 times stronger than in Fig.~\ref{fig:limits}.
 The current limit on this parameter from dinucleon decay in Super-K oxygen nuclei~\cite{nnbarSK:Abe:2011ky} is $\epsilon_{n \bar n} < 2.4 \times 10^{-24}$~eV~\cite{nnbar:Phillips:2014fgb}; thus to probe new parameter space NS temperatures $\ll 100$~K must be constrained, an unrealistic possibility with upcoming telescopes.
 
Finally, as our bounds only impact $\epsilon_{nn'} \lsim 2\times 10^{-9}~\rm eV$, set by the age of Cas A, our analysis does not exclude $n$-$n'$ oscillation explanations to the neutron lifetime anomaly. This requires $\epsilon_{nn'} \sim \Oc(10^{-7})$~eV to suppress proton production in the beam measurements of the neutron lifetime~\cite{nlifetime:Berezhiani:2018eds}. To probe this region requires extending our analysis to consider a nontrivial $n^\prime$ density which we reserve for future work~\cite{NS-MPR}.

%%%%%%%%%%%%%%%%%%%%%%%%%
\section*{Acknowledgments}
%%%%%%%%%%%%%%%%%%%%%%%%%

We thank 
Joe Bramante,
Rebecca Leane,
and
Shirley Li 
for helpful conversations on future telescopes.
The work of D.\,M. and N.\,R. is supported by the Natural Sciences and Engineering Research Council of Canada. 
T\acro{RIUMF} receives federal funding via a contribution agreement with the National Research Council Canada.
M.P. is supported in part by U.S. Department of Energy Grant No.
desc0011842.

\bibliography{refs}

\end{document}